\documentclass[aps,prl,preprint,groupedaddress]{revtex4-2}

\usepackage{graphicx}
\usepackage{dcolumn}
\usepackage{bm}

\usepackage[utf8]{inputenc}
\usepackage[T1]{fontenc}
\usepackage{mathptmx}
\usepackage{etoolbox}

\usepackage{amssymb,amsmath,placeins,epsfig,array}
\usepackage[usenames,dvipsnames]{color}
\usepackage[normalem]{ulem}
\usepackage[breaklinks,colorlinks,urlcolor=blue,citecolor=blue,linkcolor=blue]{hyperref}
\usepackage{lineno}
\usepackage{graphicx}
\usepackage{float}
\usepackage{mathptmx}

\usepackage{scalerel}
\usepackage{tikz}
\usetikzlibrary{svg.path}

\definecolor{orcidlogocol}{HTML}{A6CE39}
\tikzset{
  orcidlogo/.pic={
    \fill[orcidlogocol] svg{M256,128c0,70.7-57.3,128-128,128C57.3,256,0,198.7,0,128C0,57.3,57.3,0,128,0C198.7,0,256,57.3,256,128z};
    \fill[white] svg{M86.3,186.2H70.9V79.1h15.4v48.4V186.2z}
                 svg{M108.9,79.1h41.6c39.6,0,57,28.3,57,53.6c0,27.5-21.5,53.6-56.8,53.6h-41.8V79.1z M124.3,172.4h24.5c34.9,0,42.9-26.5,42.9-39.7c0-21.5-13.7-39.7-43.7-39.7h-23.7V172.4z}
                 svg{M88.7,56.8c0,5.5-4.5,10.1-10.1,10.1c-5.6,0-10.1-4.6-10.1-10.1c0-5.6,4.5-10.1,10.1-10.1C84.2,46.7,88.7,51.3,88.7,56.8z};
  }
}

\newcommand\orcidicon[1]{\href{https://orcid.org/#1}{\mbox{\scalerel*{
\begin{tikzpicture}[yscale=-1,transform shape]
\pic{orcidlogo};
\end{tikzpicture}
}{|}}}}

\usepackage{hyperref}
\usepackage{graphicx}
\usepackage{dcolumn}
\usepackage{bm}
\usepackage{subfigure}
\usepackage{epstopdf}
\usepackage{hyperref}
\usepackage{color}
\usepackage[utf8]{inputenc}

\makeatletter
\def\@email#1#2{%
 \endgroup
 \patchcmd{\titleblock@produce}
  {\frontmatter@RRAPformat}
  {\frontmatter@RRAPformat{\produce@RRAP{*#1\href{mailto:#2}{#2}}}\frontmatter@RRAPformat}
  {}{}
}%

\makeatother

\begin{document}

\preprint{AIP/123-QED}

\title{ \large{Spinning Hall Probe magnetic compass} }
\author {B.~Wojtsekhowski \orcidicon{0000-0002-2160-9814}} 
\email{bogdanw@jlab.org} 
\affiliation{Thomas Jefferson National Accelerator Facility{,} Newport  News{,} VA 23606} 
 
\begin{abstract}
In a large range of physics and space experiments, the direction of the magnetic field needs to be determined with an accuracy on the level of one milli radian or better.
We have proposed a new type of magnetic compass - a rotation-based one, which provides an alternating signal from the Hall probe proportional to the value of 
the magnetic field transverse to the axis of rotation.
The spinning Hall probe device has been realized. 
Alignment of the spinning axis for a minimum (zero) value of the signal allows us to find the direction of the magnetic field.
The device does not require calibration and is free of any drift problems.
The measurement of the rotation axis direction was accomplished by means of a laser and a flat mirror attached to the rotor. 
The constructed prototype achieved an accuracy for the magnetic field direction in the experiment with the polarized He-3 target
on the level of one milli radian.
\end{abstract}
\maketitle
	
\section{Introduction}

The magnetic compass for navigation at sea and on land was introduced a few thousand years ago~\cite{CMC}.
The old magnetic compass is based on a magnetized iron arrow.
It operates due to the torque between the magnetic moment of the iron and the magnetic field: $\vec B \times \vec M$.
When the vectors $\vec B$ and $\vec M$ are parallel to each other the torque is zero, so the arrow orientation is stable and 
shows the magnetic field direction.
For best precision the arrow support mechanics is optimized to have as little friction as possible.
However, accuracy much better than one degree is still hard to achieve.
Additional systematics is due to magnetic hysteresis, which leads to misalignment of the magnetic moment of the arrow and its geometrical axis.

Almost 130 years ago, A.~Righi, after discovering a huge Hall effect in Bismuth, 
built the first Hall-probe-based Earth magnetic compass, see Fig.~19 in Ref.~\cite{Campbell}.
With the advance of solid state technology, studies of the Hall probe for the magnetic compass were detailed in Ref.~\cite{Ross_1957}.
Recent progress in the field direction measurement is mostly related to the use of 3D magnetic sensors based on the Hall effect, see for example~\cite{3D}.
The angular accuracy of such a compass is very good but operation is complicated due to the electronics drift and need for 
calibration of the device as the exact orientation of the probe's plane is undefined and the probe signal drifts.
Rotation of the field detector was used widely for field mapping of the accelerator magnets and spectrometers 
with the coils, but not for precise determination of the field direction, see e.g. Ref.~\cite{Workshop-2014}.

Development of the current device was motivated by the experiments~\cite{Barabanov, wojtsekhowski_flavor_2020}  at Jefferson Lab,
where a polarized He-3 target was placed in a 25~Gauss magnetic field, see e.g. \cite{He-3}, 
the direction of which needs to be determined to 1~mrad precision.

\section{Geometrical concept of the compass}
\label{sec:concept}
\begin{figure}[ht!]
	\centering
	\includegraphics[trim = 0 0 0 0, width=.65\columnwidth, angle =0]{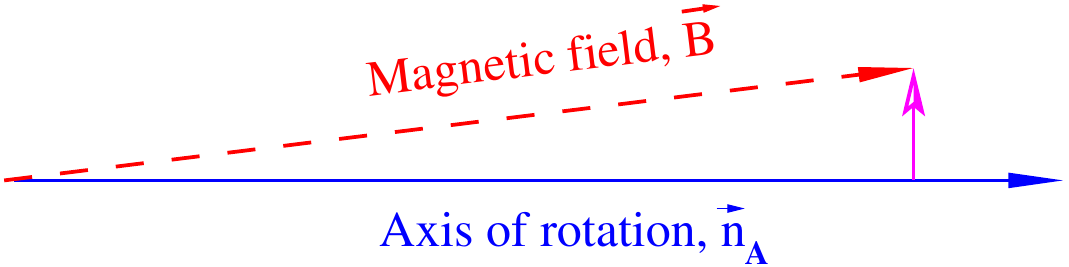}
	\caption{Diagram of the magnetic field (red) with a component (magenta) orthogonal to the compass axis of rotation (blue). The unit vector $\vec n_{_A}$ is parallel to the axis of rotation.}
\label{fig:vectors}
\end{figure}

We proposed to use the Hall probe for detection of the component of the magnetic field perpendicular to 
the axis of rotation and create an alternating signal by means of rotation,
as shown in Fig.~\ref{fig:vectors}, see also the patent application~\cite{Patent}.
The probe's plane is oriented approximately parallel to the axis of rotation.
The amplitude of the alternating signal is proportional to the magnetic field component transverse to the axis of rotation, $\vec B \times \vec n_{_A}$.
Detection of the alternating signal synchronized with the rotation phase by means of a lock-in makes the proposed device very accurate.

\begin{figure}[ht!]
	\centering
	\includegraphics[trim = 0 0 0 0, width=.8\columnwidth, angle =0 ] {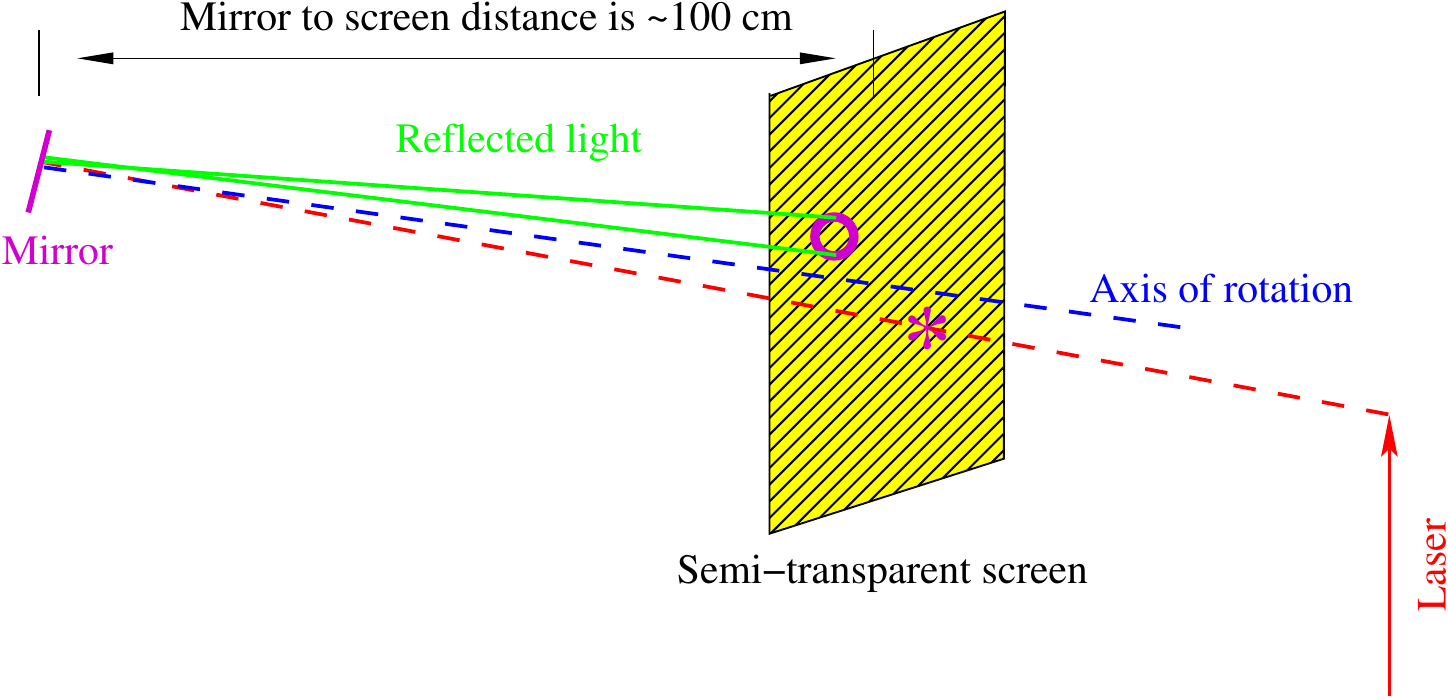}
		\includegraphics[trim = 0 20 0 20, width=.9\columnwidth, angle =0 ] {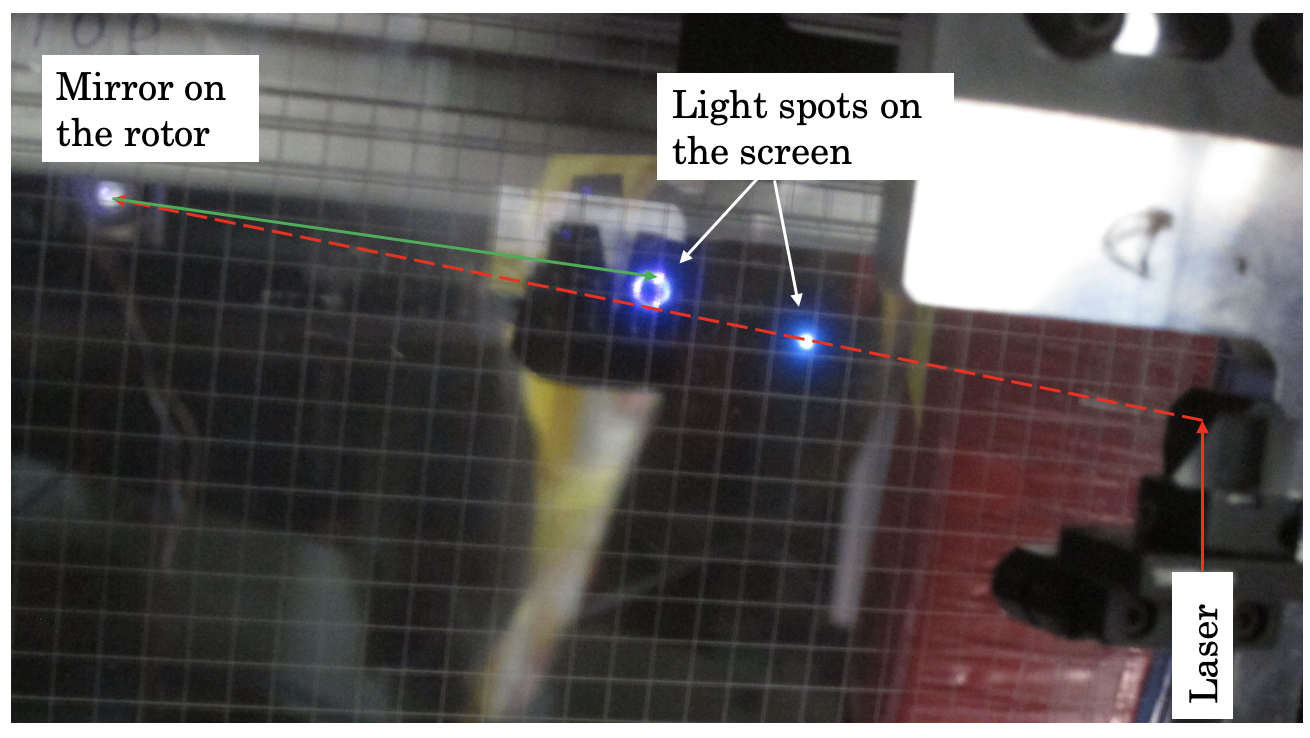}
	\caption{Diagram of optics used for determination of the rotation axis direction. 
	The coordinates of the light spot from the incident laser rays and the ring center of the light reflected from the mirror were determined by using the grid on the semi transparent screen whose location was surveyed. 
	The reflected light forms a cone due to slight non-orthogonality of the mirror plane and the rotation axis.
	A red line and a green line were added for easy comparison with the plot above.}
\label{fig:optics}
\end{figure}

The direction of the rotation axis could be determined by using the reflection of the laser beam 
from a flat mirror mounted on the rotor with the plane orthogonal to the axis of rotation, see Fig.~\ref{fig:optics}.
Mounting of the laser on the rotor provides an alternative way to determine the rotation axis direction.

\section{Components of the spinning compass}

\begin{figure}[ht!]
	\centering
	\includegraphics[trim = 0 60 0 20, width=.8 \columnwidth, angle =0 ] {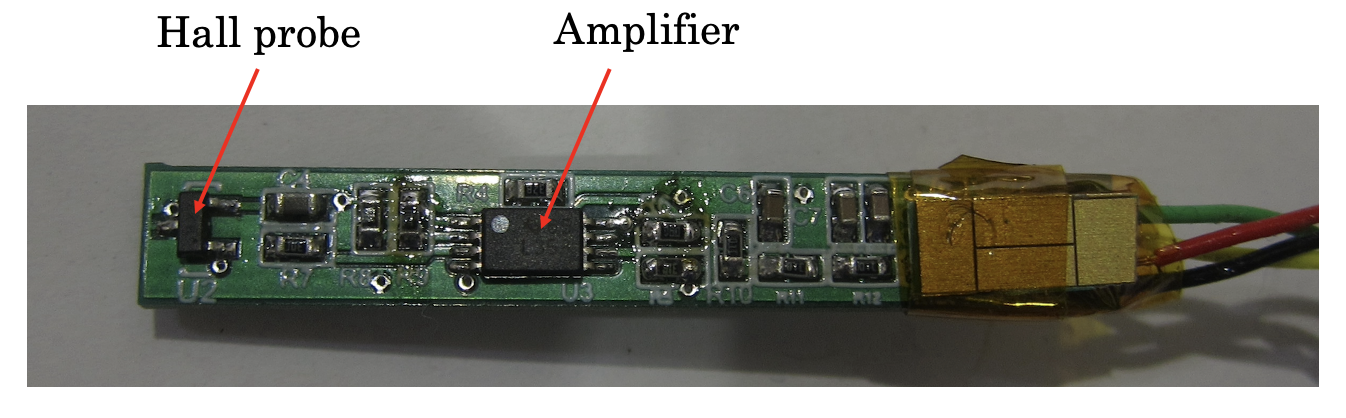}
	\caption{PC board with a Hall probe and front-end electronics.}
\label{fig:probe}
\end{figure}

Currently the Hall probe (HP) sensor has become available at low cost and its sensitivity has reached 9~mV per Gauss (after in-chip amplification), see Ref.~\cite{Probe}.
The slow instability of the output signal is resolved in the device we have proposed.
Instead of calibration of the HP we use frequent flip of the HP by means of rotation.
The probe is mounted on the PC board, see Fig.~\ref{fig:probe}, close to the axis of the rotor. 
The probe plane is approximately parallel to the axis of rotation, so the observed alternating signal is proportional to the component of the magnetic field transverse to the axis of rotation.

A two-stage amplifier allows us to increase the signal by a factor of 6, reduce sensitivity to fluctuation of the resistance of the slip-ring contacts, 
match the probe output to the high impedance of the detector (an oscilloscope in our case), and bring the frequency band closer to a desirable range ($\sim$6~Hz).
The PC board also has LED for measurement of the rotation phase.
The output signal and power to the probe and electronics are provided via a slip ring~\cite{SlipRing}.

\begin{figure}[ht!]
	\centering
	\includegraphics[trim = 0 0 0 20, width=.9\columnwidth, angle =0 ] {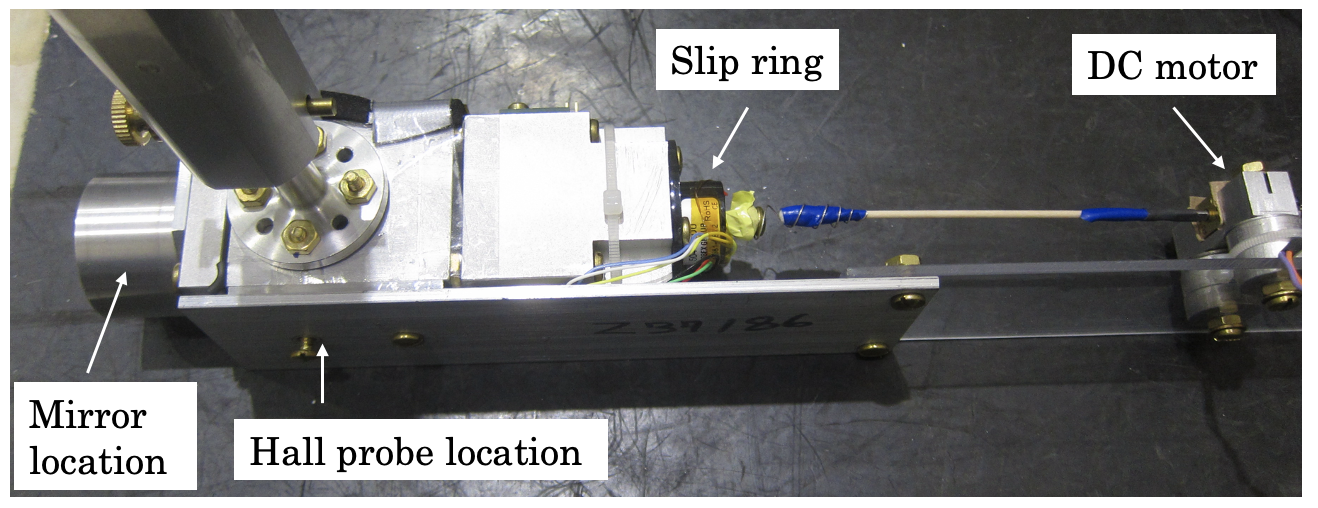}
	\caption{Spinning magnetic compass.}
\label{fig:compass}
\end{figure}

The photo resistor placed on the stationary body of the compass is used to observe the rotation phase.
The rotor is separated from the stationary body of the compass by a pair of precision ceramic bearings.
The slip ring is also attached to the stationary body, see Fig.~\ref{fig:compass}.

The optical system used for the determination of the direction of the rotation axis includes a pocket laser, a flat mirror attached to the rotor, and
two screens, one screen located between the laser and the compass and the second one behind the compass.
The screen located between the laser and the compass is semi transparent with a fine grid which allows us to measure locations of the light spots.
A picture of the system is shown in the bottom section of Fig.~\ref{fig:optics}.
The middle point between the laser incoming spot and the center of the reflected cone corresponds to the axis of rotation.
The second screen allows us to find the direction of the laser light (when the compass is removed).
Knowledge of the locations of these three light spot points on the screens is sufficient for calculation of the rotation axis direction.

Rotation of the rotor, inside which the Hall probe is located, is activated by a piezo-electrical motor~\cite{P-E_motor}, or by an air jet turbine, 
or by a DC motor~\cite{DC_motor} (the last was located at a sufficient distance $\sim$20 cm from the Hall probe along the axis of rotation).
A picture of the whole device is shown in Fig.~\ref{fig:compass}.
	
\section{Measurement of the field direction}
	
\begin{figure}[ht!]
	\centering
	\includegraphics[trim = 0 0 0 0, width=0.8\columnwidth, angle =0 ] {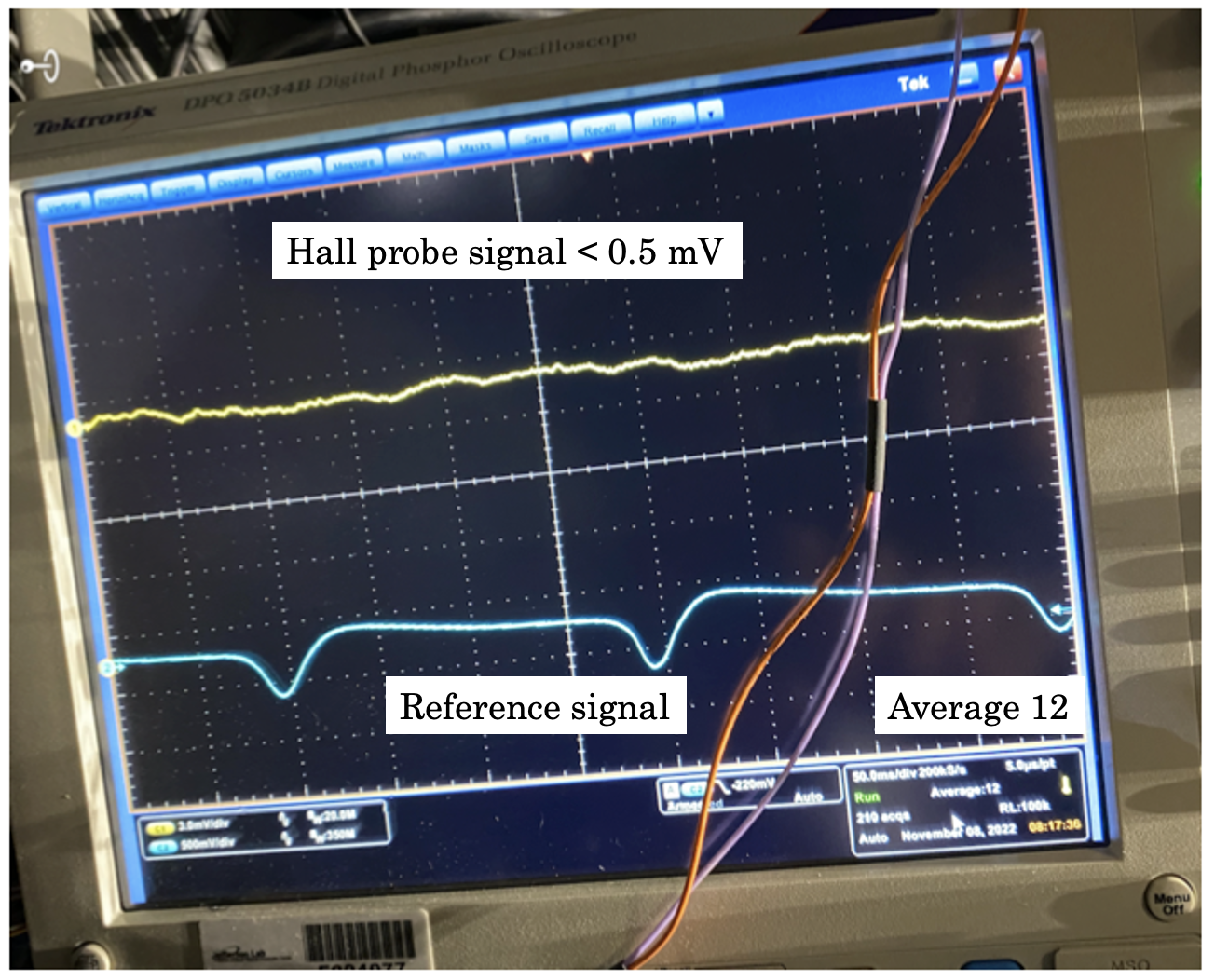}
	\caption{Signal from the compass aligned to the magnetic field.}
\label{fig:aligned}
\end{figure}

During the measurement the compass was mounted in the center of the target on a support which allows adjustment of the device orientation.
The fine tuning of the orientation was done doing minimization of the Hall probe signal (averaged for 12 triggers) on the oscilloscope, see Fig.~\ref{fig:aligned}.

The next step is a picture taken of the semi transparent screen, see Fig.\ref{fig:optics}.
After that the compass was shifted along the target to allow the laser light to reach 
a second screen (separated from the first one by 200 cm) and a picture was taken
from which the coordinates of the light spot were determined with accuracy of 1~mm.
These three locations (two on the semi transparent screen and one on the second screen) 
allow determination of the direction of the axis of rotation.

For a check of the systematics, the constructed device was placed inside the area protected 
by the multiple $\mu$-metal screens.
The observed amplitude of the alternating signal corresponds to the field of 1 milli Gauss.
For the case of He-3 target this level of systematics corresponds to angular accuracy on the level of 0.04 mrad.

\section{Conclusion}

Rotation of the Hall probe allows us to create an alternating signal and do high precision determination of the magnetic field direction.
We realized the spinning Hall probe device and reached a precision for the magnetic field direction on the level of 1 mrad for the case of a 25 Gauss field 
with only an average of 12 shots using the oscilloscope (corresponding to 6 seconds of data taking).
With the use of a lock-in detector, the limit could be advanced at least by a factor of 10.
The device could find application in many fields, including air and sea navigation, space exploration, and physics experiments.

\section{Acknowledgement}	

We would like to thanks to M.~Jones, T.~Keppel, A.~Weisenberger, and R.~Wines for their interest and support of the project.
We extend gratitude towards J.~Armstrong, C.~Cuevas, C.~Gould, W.~Henry, I.~Rachek, D.~Spiers, A.~Tadepalli, and J.~Wilson 
for collaboration in the compass development.
This work was supported in part by the US DOE Office of Science and Office of Nuclear Physics under the contracts DE-AC02-05CH11231, 
DE-AC02-06CH11357, and DE-SC0016577, as well as DOE contract DE-AC05-06OR23177, under which JSA, LLC operates JLab.

\bibliography{Compass_arXiv.bib}

\end{document}